# *Correlating Raman Spectral Signatures with Carrier Mobility in Epitaxial Graphene: A Guide to Achieving High Mobility on the Wafer Scale*

Joshua A. Robinson,[1,2] Maxwell Wetherington,[1] Joseph L. Tedesco,[3] Paul M. Campbell,[3] Xiaojun Weng,[1,2] Joseph Stitt,[2] Mark A. Fanton,[1] Eric Frantz,[1] David Snyder,[1] Brenda L. VanMil,[3] Glenn G. Jernigan,[3] Rachael L. Myers-Ward,[3] Charles R. Eddy, Jr.,[3] and D. Kurt Gaskill[3]

1. Electro-Optics Center
   The Pennsylvania State University
   559A Freeport Rd., Freeport, PA 16229
2. The Materials Research Institute
   The Pennsylvania State University
   University Park, PA, 16802
3. Naval Research Laboratory
   4555 Overlook Ave., Washington, D.C. 20375

**We report a direct correlation between carrier mobility and Raman topography of epitaxial graphene (EG) grown on silicon carbide (SiC). We show the Hall mobility of material on the Si-face of SiC [SiC(0001)] is not only highly dependent on thickness uniformity but also on monolayer strain uniformity. Only when both thickness and strain are uniform over a significant fraction (> 40%) of the device active area does the mobility exceed 1000 $cm^2$/V-s. Additionally, we achieve high mobility epitaxial graphene (18,100 $cm^2$/V-s at room temperature) on the C-face of SiC [SiC(000-1)] and show that carrier mobility depends strongly on the graphene layer stacking. These findings provide a means to rapidly estimate carrier mobility and provide a guide to achieve very high mobility in epitaxial graphene. Our results suggest that ultra-high mobilities (>50,000 $cm^2$/V-s) are achievable via the controlled formation of uniform, rotationally faulted epitaxial graphene.**

The recent success of graphene transistor operation in the giga-hertz range has solidified the potential of this material for high speed electronic applications.[1,2] Realization of graphene technologies at commercial scales, however, necessitates large-area graphene production, as well as the ability to rapidly characterize its structural and electronic quality. Graphene films can be produced by mechanical exfoliation from bulk graphite,[3,4] reduction of graphite-oxide,[5,6] chemical vapor deposition on catalytic films,[7] or via Si-sublimation from bulk SiC substrates.[8-12] The last technique currently appears to hold the most promise for large-area electronic grade graphene, and already shows tremendous potential for high-frequency device technologies.[2] Nevertheless, precise control of the graphene electronic properties (*i.e.* mobility) over large areas is necessary to enable graphene-based technological applications. Realization of such control will come through an intimate understanding of the process-property-performance relationship and the role that graphene thickness, strain, and layer stacking plays in this relationship over very large areas up to full wafers. Of the characterization techniques used for layer thickness determination,[13-19] Raman spectroscopy is arguably the simplest and fastest, especially for exploring monolayer EG on SiC(0001) (referred to as $EG_{Si}$) and EG layer stacking on SiC(000-1) (referred to as $EG_c$).[15-19]

Characterization of EG via Raman spectroscopy requires fitting the 2D Raman peak.[15,16,20] Raman spectra of $EG_{Si}$ fit by one or four Lorentzian functions are characteristic of monolayer or bilayer graphene, respectively.[15] Figure 1a demonstrates layer thickness evaluation for monolayer and bilayer $EG_{Si}$ via Lorentzian fitting of the 2D Raman spectra. To further validate these thickness measurements, cross-sectional transmission electron microscopy (TEM) was performed (Fig. 1b,c). The TEM micrographs in Fig.1b,c include a transition layer (Layer 0), which is in direct contact with the SiC substrate, and generally is not considered graphene.[21,22] The subsequent layers above Layer 0 constitute the electrically active graphene, and give rise to its unique properties. As such, these first two upper layers are considered monolayer (Fig. 1b) and bilayer (Fig.1c) graphene, respectively. In contrast, $EG_C$ is generally several layers thick, but can exhibit a Raman signature similar to both monolayer $EG_{Si}$ and bulk graphite. Figure 1d displays typical 2D Raman spectra from $EG_C$ that are fit to one or two Lorentzians. This fitting yields information on layer stacking of the multilayer films, and is discussed in detail later.[23]

Two-dimensional mapping of EG via Raman spectroscopy, termed Raman topography, provides a means to explore how thickness and strain uniformity influence carrier mobility. For the purpose of characterizing the relationship between $EG_{Si}$ uniformity and carrier mobility, it is not necessary to distinguish between thickness uniformity and strain uniformity. The presence of non-uniformities in either case detrimentally affects EG carrier mobility. As a result, the 2D Raman peak position is a sufficient metric to monitor film uniformity. High mobility $EG_{Si}$ (> 1000 $cm^2$/V-s) exhibits a very uniform Raman signature. Compressively strained (0.9 – 1.0%, based on Ferralis, et al.[24]) monolayer $EG_{Si}$ is present across the SiC terrace centers, with bilayer graphene only at the terrace edges (Fig. 2a). In contrast, Raman topography of low mobility, monolayer $EG_{Si}$ (Fig. 2c) exhibits significant variation in the 2D peak position (2690 – 2760 $cm^{-1}$); which indicates that, even with uniform thickness, a high density of transitions from 0.1 – 1.1% compressive strain exist.[16,24] High densities of small domains in the Raman topography map are well correlated with low carrier mobility $EG_{Si}$ (Fig.2e), which presumably originates from carrier scattering at the domain boundaries. As the Raman topography domain size increases, the carrier mobility improves to 600 $cm^2$/V-s; however, achieving mobilities > 1000 $cm^2$/V-s requires that the Raman topography map be uniform over > 50% of the device. Ultimately, to increase carrier mobility > 2000 $cm^2$/V-s, $EG_{Si}$ uniformity must be present over the entire device (Fig. 2e). We note that we have not explicitly explored the effects of strain alone on the mobility of $EG_{Si}$, but suspect that it plays a key role.

$EG_C$ exhibits superior carrier mobility compared to $EG_{Si}$.[25,26] However, thickness determination of $EG_C$ by Raman spectroscopy is difficult because it consistently grows > 3 layers thick, resulting in a 2D Raman peak that is generally fit using two Lorentzian functions and thus is similar to bulk graphite. Equally important to layer thickness is layer-stacking, which can also be extracted from the 2D Raman spectra.[17,18,19,23] Multilayer $EG_C$ films can exhibit a 2D Raman peak characteristic of turbostratic graphite;[17] however, the layer-stacking may not be completely random. Evidence suggests a mixture of graphene layers rotated by 30° or $\pm2°$ (R30 or $R2^{\pm}$) with respect to the SiC substrate exist (typically described as R30/$R2^{\pm}$ fault pairs).[23] Moreover, $EG_C$ with R30/$R2^{\pm}$ fault pairs contain significantly larger grains (>10x),[23] smaller 2D Raman peak widths (< 30 $cm^{-1}$ compared to > 40 $cm^{-1}$), and the absence of the defect-induced Raman peak (D-peak).[27,28] Based on these results, we identify multilayer $EG_C$ exhibiting a narrow 2D Raman peak and fit by a single Lorentzian (Fig.1d, top) as rotationally faulted epitaxial graphene ($EG_{RF}$). In contrast, graphene layers exhibiting wider 2D Raman peaks resemble those of bulk graphite, which has AB or Bernal stacking (Fig.1d, middle and bottom). We therefore identify EG with 2D peaks in this range as Bernal stacked EG.

We find the stacking order of $EG_C$ strongly affects the Hall carrier mobility. Correlation with Raman topography enables identification of very high mobility $EG_C$ without electrical measurements. While the

EG$_C$ 2D peak position varies by only 8 cm$^{-1}$ (2716 – 2724 cm$^{-1}$) (Fig. 1d), the 2D peak full width at half maximum (FWHM) ranges between 25 cm$^{-1}$ and 75 cm$^{-1}$, which we attribute to a significant variation in the layer-stacking of EG$_C$. As the Lorentzian components increase in separation, the mobility decreases significantly from a high 18,100 cm$^2$/V-s to a low of 50 cm$^2$/V-s.[29] The increase in the 2D peak width from a narrow single Lorentzian to a broad two Lorentzian fit indicates EG$_C$ is subject to two types of stacking: EG$_{RF}$ and Bernal stacked. Rotationally faulted EG$_C$ exhibits very high carrier mobility due to reduced phonon scattering, which is consistent with this type of graphene layer-stacking.[17,23,23] We note that Hall crosses fabricated on EG$_C$ consist of both uniform (FWHM < 10% variation, Fig. 3b) and non-uniform (FWHM > 10% variation, Fig. 3c) EG$_C$, and that the correlation between FWHM and mobility only exists for those devices with uniform EG$_C$ films (Fig. 3a). Non-uniform EG$_C$ shows no such dependence.

The formation of uniform, high mobility graphene is paramount to the success of graphene-based technologies. We present here the first systematic study to link effects of graphene thickness, strain, and structure on carrier mobility. We have investigated EG$_{Si}$ and EG$_C$ and have correlated carrier mobility with Raman spectral signatures. This provides a means for rapid, non-destructive evaluation of epitaxial graphene carrier mobility. It should therefore be possible to determine (prior to device fabrication) by Raman topography and analysis of the 2D peak shape whether a film of epitaxial graphene will have high or low carrier mobility. The thickness of EG$_{Si}$ is controllable down to a single monolayer; however, we have shown that mastering control over both thickness and strain uniformity is essential to achieve high mobility EG$_{Si}$. Through careful study of the preparation and synthesis parameters to control strain uniformity in the material, one could achieve large graphene domains that yield mobilities significantly higher than 2000 cm$^2$/V-s on the Si-face of SiC. Additionally, high room-temperature mobility (>18,000 cm$^2$/V-s) epitaxial graphene has been achieved on EG$_C$. We have shown that carrier mobility of EG$_C$ is strongly correlated with stacking order of the graphene layers. High mobility graphene is only realized when rotationally faulted epitaxial graphene (EG$_{RF}$) is present, represented by a narrow 2D Raman spectra fit by a single Lorentzian. Extrapolation of the data in Figure 3 suggests that the realization of ultra-high mobility EG$_c$ is possible via the development of defect-free EG$_{RF}$ layers on SiC(000-1) exhibiting narrow 2D Raman spectra (e.g.,2D line widths below 23 cm$^{-1}$ should yield mobilities >50000 cm$^2$/Vs).

## *Methods Summary*

Epitaxial graphene films EG$_{Si}$ and EG$_C$ were grown on the (0001) and (000-1) faces (Si- and C-face, respectively) of on-axis, 16x16 mm, semi-insulating 4H- and 6H-SiC substrates, using a commercial Aixtron/Epigress VP508 Hot-Wall CVD reactor.[12] This system is capable of accommodating up to 76.2 mm SiC wafers. Mobility measurements of epitaxial graphene were performed using Van der Pauw test structures (Hall crosses) with widths of 10 µm for EG$_{Si}$, and 10 or 2 µm for EG$_C$. A WITec confocal Raman microscope (CRM) with a 488 nm laser wavelength, diffraction limited lateral resolution of ~ 340 nm, and spectral resolution of 0.24 cm$^{-1}$ was utilized for Raman spectroscopy. Graphene films on both SiC(0001) and SiC(000-1) experience variation in the 2D Raman peak. Similar to surface topographic variation measurements in atomic force microscopy, Raman topography is used to identify the length scale over which variations in the spectral signature occurs. Further detail is provided in the online supplementary information.

**Supplementary Information** is linked to the online version of the paper at www.nature.com/nature.

**Acknowledgements** The authors would like to acknowledge funding support through The Penn State Electro-Optics Center IRAD 01830.70, and The Naval Research Laboratory Nanoscience Institute. Additionally, support for the WiteC Raman system was provided by the National Nanotechnology Infrastructure Network at Penn State. J.L.T. and B.L.V. also acknowledge support from the American Society for Engineering Education for postdoctoral fellowships.

**Author Information** Reprints and permissions information is available at npg.nature.com/reprintsandpermissions. The authors declare no competing financial interests. Correspondence and requests for materials should be addressed to J.R. (jrobinson@psu.edu).

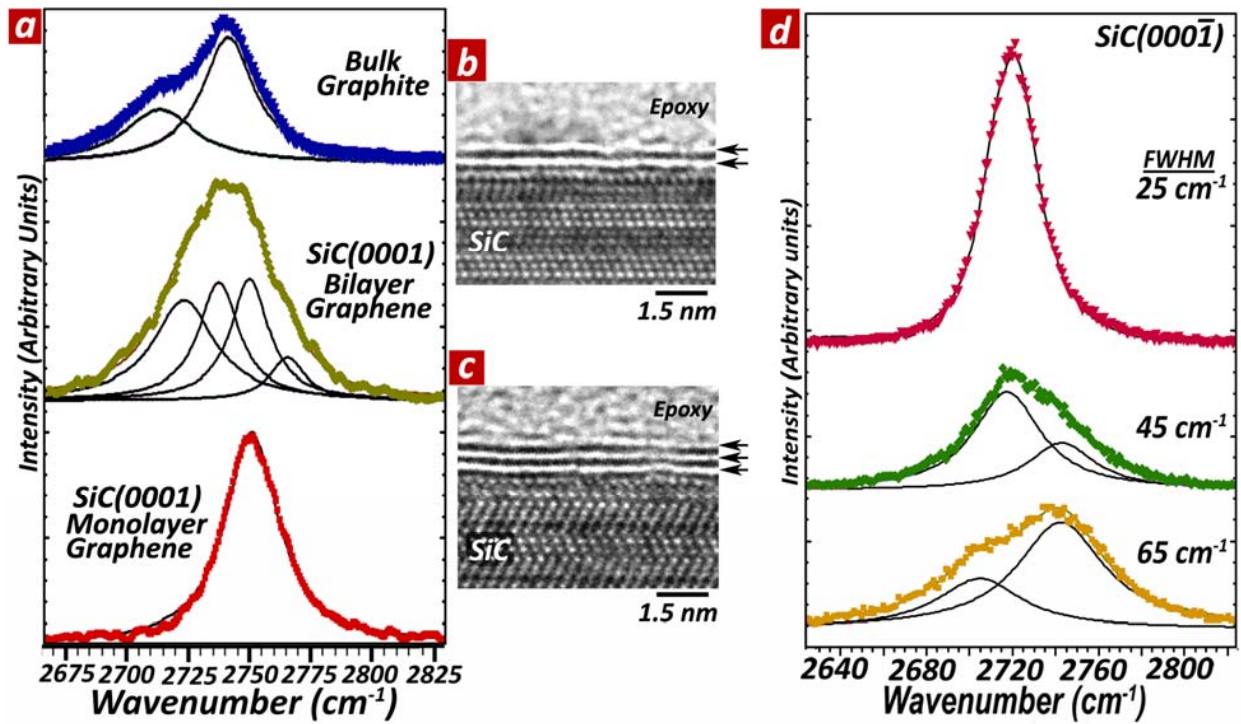

**Figure 1: The 2D Raman peak of epitaxial graphene is used to rapidly identify (a) monolayer and bilayer graphene on SiC(0001) and layer stacking on SiC(000-1).** Film thickness measurement are confirmed via TEM (b,c). Transmission electron micrographs show the SiC/graphene transition layer,[21,22] as well as additional layers which make up the electronic and structural properties of monolayer (b) and bilayer (c) graphene. Graphene layer stacking is correlated with 2D Raman peak width for EG on SiC(000-1)(d). Those graphene films exhibiting narrow (< 30 cm$^{-1}$) peak widths are fit to a single Lorentzian, suggesting the presence of rotationally faulted EG$_C$;[17,23] while peaks with FWHM > 35 cm$^{-1}$ are similar in shape to that of bulk graphite consisting of an AB-type stacking order.

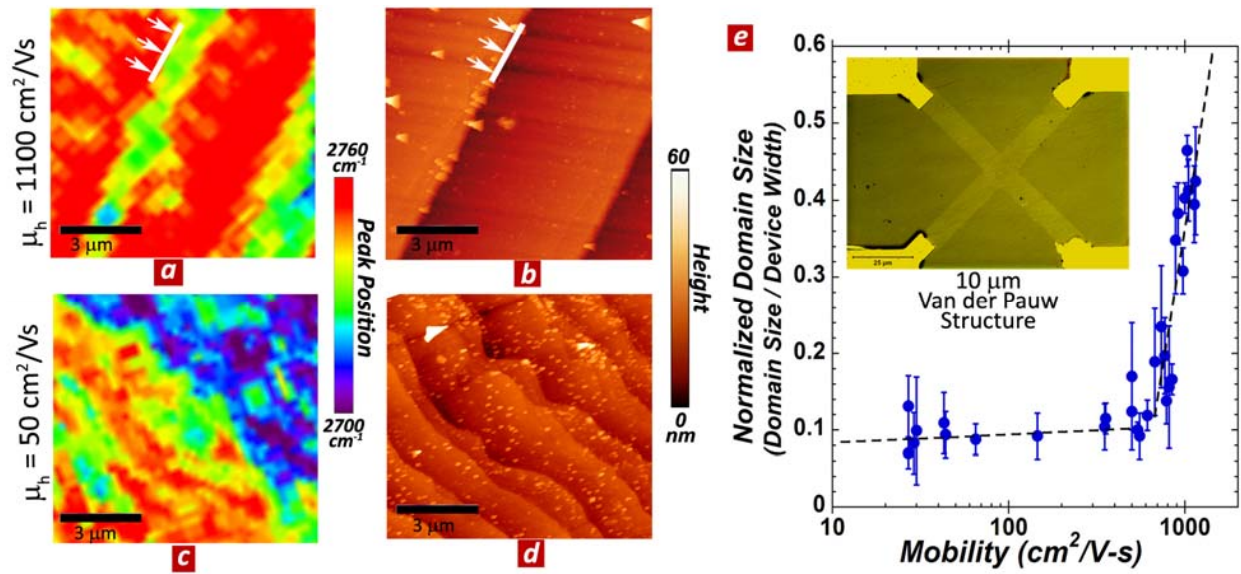

**Figure 2: Raman topography (a,c), atomic force microscopy (b,d), and Hall mobility (e) data are used to identify the influence of graphene uniformity on carrier mobility of EG$_{Si}$.** High mobility EG$_{Si}$ exhibits uniformly strained graphene with minimal thickness variation, which is identified in Raman topography by a uniformly distributed 2D Raman peak position. Low mobility graphene (c,d) can be mono- or bilayer, however, the length scale of the Raman topography uniformity is significantly smaller than the device length scale. The normalized Raman topography domain size (domain size/device width) and Hall mobility from thirty EG$_{Si}$ Hall crosses indicates that the film uniformity significantly influences carrier mobility (e). Arrows in (a,b) indicate the location of the SiC terrace step edge. The dashed line in (e) is presented as a guide for the reader.

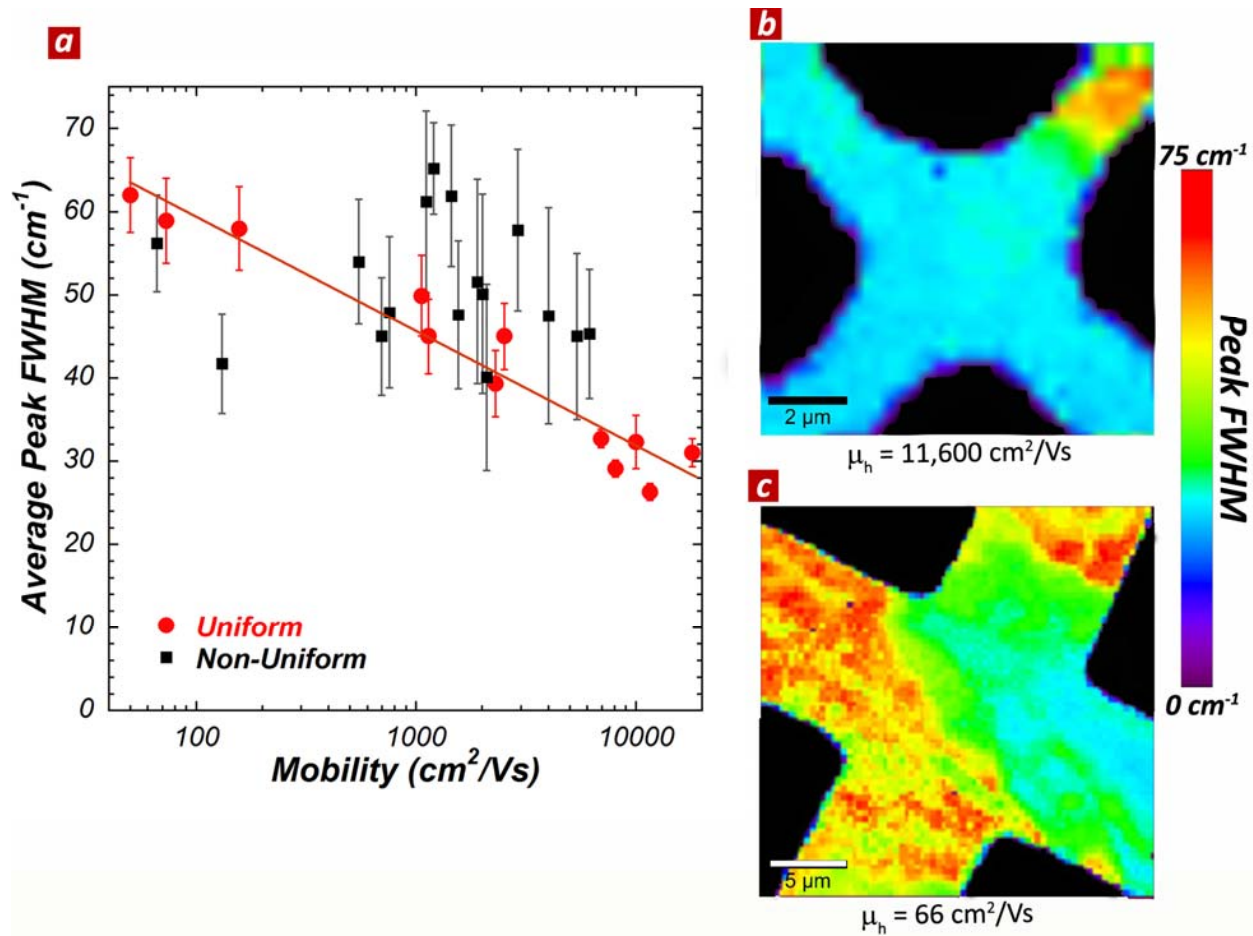

**Figure 3: The Raman 2D peak width is strongly correlated with EG$_C$ carrier mobility.** The peak width of the 2D Raman spectra, which describes the stacking order of the graphene layers, is strongly correlated with graphene carrier mobility (a) for Hall crosses exhibiting uniform EG$_C$ (b). In contrast, drawing a correlation between mobility and peak FWHM is not possible for those devices with non-uniform Raman peak widths (a,c). The values presented in (a) are average peak width values of all spectra taken at the center of each Hall cross (b,c), and consist of > 100 spectra each.

*Correlating Raman Spectral Signatures with Carrier Mobility in Epitaxial Graphene: A Guide to Achieving High Mobility on the Wafer Scale*

Joshua A. Robinson,[1,2] Maxwell Wetherington,[1] Joseph L. Tedesco,[3] Paul M. Campbell,[3] Xiaojun Weng,[1,2] Joseph Stitt,[2] Mark A. Fanton,[1] Eric Frantz,[1] David Snyder,[1] Brenda L. VanMil,[3] Glenn G. Jernigan,[3] Rachael L. Myers-Ward,[3] Charles R. Eddy, Jr.,[3] and D. Kurt Gaskill [3]

1. Electro-Optics Center
   The Pennsylvania State University
   559A Freeport Rd., Freeport, PA 16229
2. The Materials Research Institute
   The Pennsylvania State University
   University Park, PA, 16802
3. Naval Research Laboratory
   4555 Overlook Ave., Washington, D.C. 20375

## *Methods*

### *Graphene Synthesis and Hall Mobility Evaluation*

Graphene films are epitaxially grown on the (0001) and (000-1) faces (Si- and C-face, respectively) of on-axis, semi-insulating 4H- and 6H-SiC substrates using a commercial Aixtron/Epigress VP508 Hot-Wall CVD reactor.[1] Graphene on SiC(0001) is referred to as $EG_{Si}$, graphene on SiC(000-1) is referred to as $EG_C$. Synthesis of epitaxial graphene is performed between 1225 and 1700°C for 30 to 90 min at $10^{-6}$ to $10^{-4}$ mbar. Graphene Van der Pauw structures were fabricated using standard photolithography to define the Hall crosses with 10μm and 2μm wide central regions. Titanium/gold bilayers were used for ohmic contacts. Mobility measurements of epitaxial graphene were performed using Van der Pauw structures (Hall crosses) with widths of 10 μm for $EG_{Si}$, and 10 or 2 μm for $EG_C$. Mobility values were acquired at 300 K using a permanent magnet Hall system.

### *WiTec Confocal Raman Microscopy*

A WITec confocal Raman microscope with a 488 nm laser wavelength, diffraction limited lateral resolution of ~ 340 nm, and spectral resolution of 0.24 $cm^{-1}$ was utilized for Raman spectroscopy. Raman spectra were collected using a spatial step size of 300 - 500nm, laser power of ~60mW (at sample), with an integration time of 0.5 – 1 s. Using the 2D peak of graphene eliminates the need for deconvolution of the SiC/Graphene spectra because SiC does not exhibit a Raman signature in this spectral range. As a result, we chose to monitor the 2D peak of graphene as a means of rapid identification of the net effect of thickness and strain on $EG_{Si}$, and layer stacking on $EG_C$. The position of each Raman peak was identified using the center of mass of the peak over the spectral range of 2600 – 2900 $cm^{-1}$.

Layer thickness of $EG_{Si}$ in this work is identified via peak shape in the following manner: 1) each spectrum is normalized based on maximum peak intensity; 2) peaks are fit to one, two, or four Lorentzian functions. Spectra that can be fit to a single Lorentzian are considered to be monolayer graphene whereas those fit to the sum of four Lorentzians are bilayer graphene.[2,3] Peaks fit using two Lorentzian functions are categorized as a transition layer (T), which are further categorized as a thickness transition, or a strain state transition of monolayer graphene.[16] Because $EG_C$ is much thicker than $EG_{Si}$, atomic force microscopy (AFM) is utilized to measure $EG_C$ thickness. Layer stacking in $EG_C$ and $EG_{Si}$ is identified using the 2D peak full width at half maximum, and component fitting. Systematic stacking of multi-layer graphene is found exclusively in $EG_{Si}$, while $EG_C$ can exhibit a range of stacking order.

*Raman Topography*

Raman topography maps typically included a 20x20 µm region at the center of each Hall cross, and consisted of 1600-3600 spectra in each map. Both $EG_{Si}$ and $EG_C$ films experience variation in 2D Raman peak position, width, and shape. Similar to surface topographic variation measurements in AFM, Raman topography ($R_T$) is used to identify the length scale over which variations in the spectral signature occurs. The Raman topography domain size of $EG_{Si}$ is defined as the diameter (width) of those regions in a $R_T$ map with similar peak positions (deviation < 10% around mean). Alternatively, $R_T$ domain size of $EG_C$ is defined as the diameter (width) of those regions in a $R_T$ map with common peak widths (i.e. width = 25 $\pm$ 2 cm$^{-1}$). Significant non-uniformity in the 2D Raman peak position or width of the graphene in a Hall cross (standard deviation > 10%) results in these devices being defined as "non-uniform."

*Strain Calculations in Monolayer Epitaxial Graphene on SiC(0001)*

Hydrostatic compressive strain in monolayer $EG_{Si}$ is calculated according to Ferralis *et al.*:[4]

$$\frac{\Delta\omega}{\omega_o} = -\gamma_m * tr(\varepsilon_{ij}) \quad \textbf{(Eq. 1)}$$

where $\omega_o$ is the peak position of strain free (exfoliated) graphene, $\Delta\omega$ is the shift above $\omega_o$ due to strain, $\gamma m$ is the Gruniesen parameter, and $tr(\varepsilon_{ij})$ is the trace of the strain. For uniformly strained graphene (x and y strain), $tr(\varepsilon_{ij})$ = strain. Table 1 provides data for both the G and 2D peaks of graphene.

**Table 1:** Parameters used to calculate hydrostatic compressive strain in epitaxial graphene.

| | |
|---|---|
| $\omega_o^{2D}$ | 2685 cm$^{-1}$ |
| $\omega_o^{G}$ | 1585 cm$^{-1}$ |
| $\gamma_{2D}$ | 2.7 |
| $\gamma_G$ | 1.8 |

## Raman Topography of SiC(0001) Graphene Hall Crosses

Raman topography of $EG_{Si}$ yields information on thickness and strain uniformity. We present additional Raman topographic maps of $EG_{Si}$ in Fig. S1, with corresponding room temperature Hall effect mobility values.

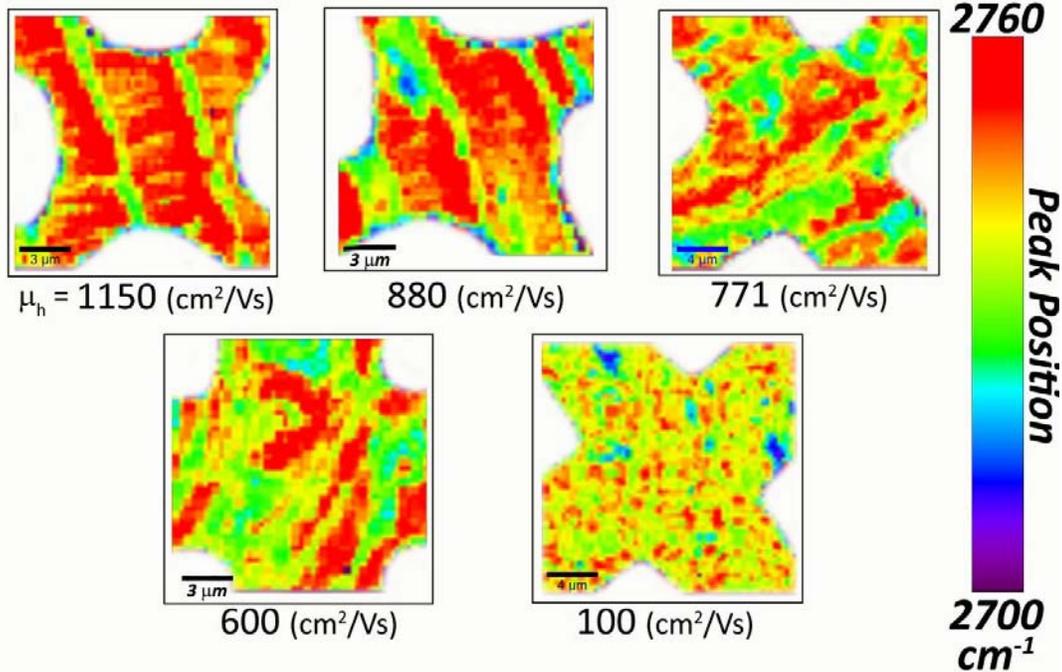

**Figure S1: Raman topographic maps showing the 2D Raman peak center-of-mass.** Five (5) Hall crosses mapped by Raman topography illustrate the influence of graphene uniformity on carrier mobility. Note that as the thickness/strain domain size decreases from >50% (top left) to ~5% (bottom right) of the device width, the mobility is reduced by an order of magnitude.

## Raman Topography of SiC(000-1) Graphene Hall Crosses

Raman topography of $EG_C$ has shown that layer stacking of $EG_C$ is a key metric for rapidly identifying high mobility epitaxial graphene on the C-face of SiC (Fig. S2). There is a distinct lack of layer stacking order in high mobility $EG_C$, described as rotationally-faulted $EG_C$ ($EG_{RF}$), which yields graphene layers that are electronically isolated from each other. Figure S2 represents a series of Hall crosses with varying peak widths. $EG_C$ with narrow peak widths (FWHM < 32 cm$^{-1}$) exhibit a room temperature mobility > 5,000 cm$^2$/V-s, while those devices exhibiting ordered stacking of the graphene layers (FWHM > 35cm$^{-1}$) exhibit a room temperature Hall mobility < 3,000cm$^2$/V-s. Additionally, the correlation between Raman 2D peak width and mobility is significantly stronger than with $EG_C$ film thickness. Figure S3 is a comparison of uniform $EG_C$ films measured via Raman topography and atomic force microscopy. While the $EG_C$ thickness varies significantly, the width of the 2D Raman peak of $EG_C$ is well correlated with mobility.

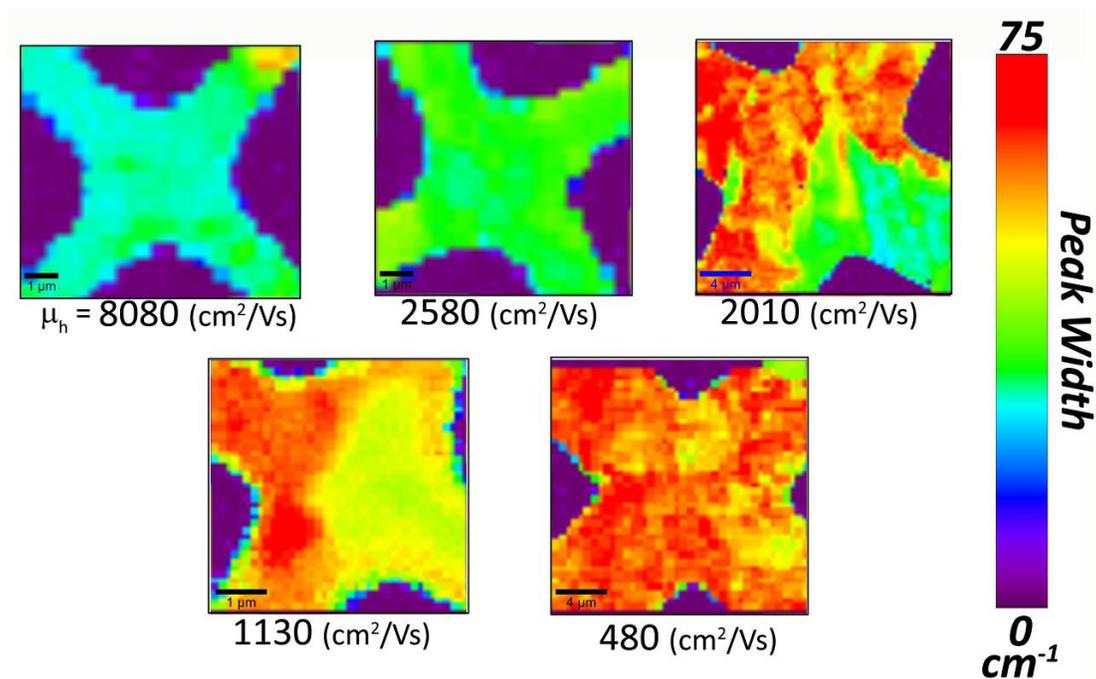

**Figure S2: Raman topography of EG$_C$ is a strong indicator of the Hall carrier mobility.** Raman topography maps of devices ranging from 480 to 8,080 cm$^2$/Vs indicate that as the 2D Raman peak width increases from < 30cm$^{-1}$ to > 50cm$^{-1}$, the Hall carrier mobility degrades by > 20x.

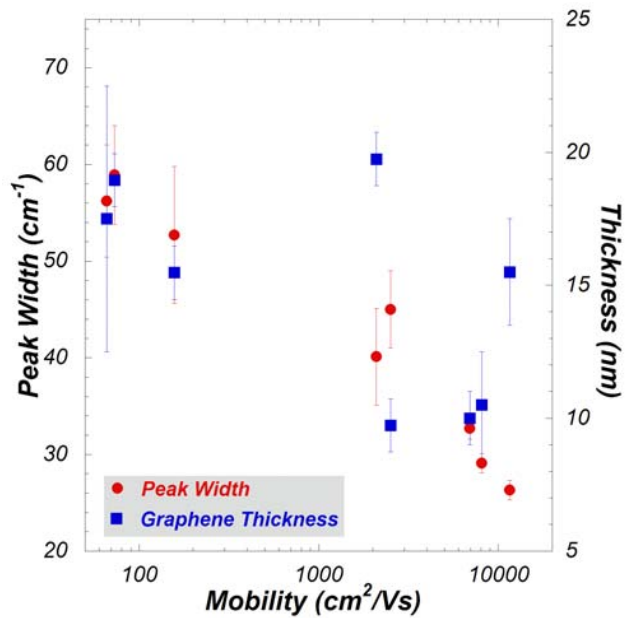

**Figure S3: Comparison of Raman 2D peak width and EG$_C$ film thickness with mobility.** Note that while thickness of the EG$_C$ is not well correlated, peak width (layer stacking and in-plane perfection) is well correlated with mobility.